\documentclass[preprint,showpacs,preprintnumbers,amsmath,amssymb]{revtex4}

\usepackage{graphicx}
\usepackage{dcolumn}
\usepackage{bm}

\newcommand{\beq}{\begin{equation}}
\newcommand{\eeq}{\end{equation}}

\begin{document}

\title{A Black Hole Life Preserver}

\author{J. Richard Gott}
  \email{jrg@astro.princeton.edu}
  \affiliation{Department of
  Astrophysical Sciences, Princeton University, Princeton, NJ 08544}
\author{Deborah L. Freedman}
  \affiliation{Harvard University Department of Astronomy,
  60 Garden St., MS-10, Cambridge, MA 02138}

\date{August 19, 2003}

\begin{abstract}
Since no one lives forever, all a life preserver can really do is prolong 
life for longer than would have otherwise been the case.  With this rather 
limited definition in mind we explore in this paper whether in principle you 
can take a life preserver with you to protect you (for a while at least) 
against the tidal forces encountered on a trip inside a black hole.
\end{abstract}

\pacs{04.70.-s }

\maketitle

It is well advertised that if you fall into a non-rotating Schwarzschild
 black hole you will be torn apart by tidal forces.  As you free-fall in on a 
radial trajectory you will experience tidal acceleration given by 
$a_z = (2M/r^3) z, a_x = - (M/r^3) x, a_y = - (M/r^3) y$, 
where, using geometrized units, $M$ is the mass of the black hole, $r$ is the 
circumferential radius appearing in the Schwarzschild metric, and $x, y, z, t$
 is a local freely falling coordinate system.  If you fall in feet-first there
 will be a differential tidal acceleration between your head and feet of 
$a_{net} = (2M/r^3)h$, where $h = 1.8$ meters is your 
height.  It is like being stretched on a rack (by $a_z$) and simultaneously 
crushed in an iron maiden (by $a_x$ and $a_y$).  You can stand this up to 
$a_{crit}$ = 10g = 98 m s$^{-2}$.  Fighter pilots withstand 9g turns for 
example.  But beyond 10g's, the tidal acceleration will cause pain 
and dismemberment.  Thus, the tidal acceleration becomes torturous at 
$r_c = (2Mh/a_{crit})^{1/3}$.  (You are still comfortable as you enter the 
event horizon of the hole only if $M > 13,800$ solar masses, so $r_c < 2M$.) 
 What is not generally recognized is that the period of torture is relatively 
brief.  Suppose you fall into the black hole from a large distance (i.e., 
zero kinetic energy at infinity).  The proper time for you to fall from radius
 $r_c$ to $r = 0$ is 
$t_c = (4M/3)(r_c/2M)^{3/2} = (2/3)(h/a_{crit})^{1/2}  = 0.0904$ sec, 
independent of the mass of the black hole.  Can you lower $t_c$, so that you
 live longer and are tortured for less time? 	

You can if you take along a life preserver: a ring of mass $M_R$ and radius 
$R$ (see Figure 1).  It even looks like a life preserver!  The ring produces 
to first order near the origin $(x, y, z \ll R)$ a tidal acceleration
 $a_z = - (M_R/R^3) z, a_x = (M_R/2R^3) x, a_y = (M_R/2R^3) y$, 
where $M_R \ll M$, $R \ll r$, and $M_R \ll R$ so that the ring's 
gravitational field is approximately Newtonian.  If $M_R/R^3 = 2M/r^3$, the 
tidal acceleration by the ring exactly cancels the tidal acceleration by the 
black hole.  If we adjust the forces in the ring so that it is just neutrally
 stable against collapse due to its own self-gravity, each piece of the ring
will fall radially inward keeping $R \propto r$ during the infall 
so the tidal force of the black hole is countered all the way in.  Of course, 
this eventually fails when $R$ becomes smaller than your waist or, more
 precisely, when $R$ is no longer much greater than $h$.	

\begin{figure}[h]
\vspace{1cm}
\includegraphics[width=11cm]{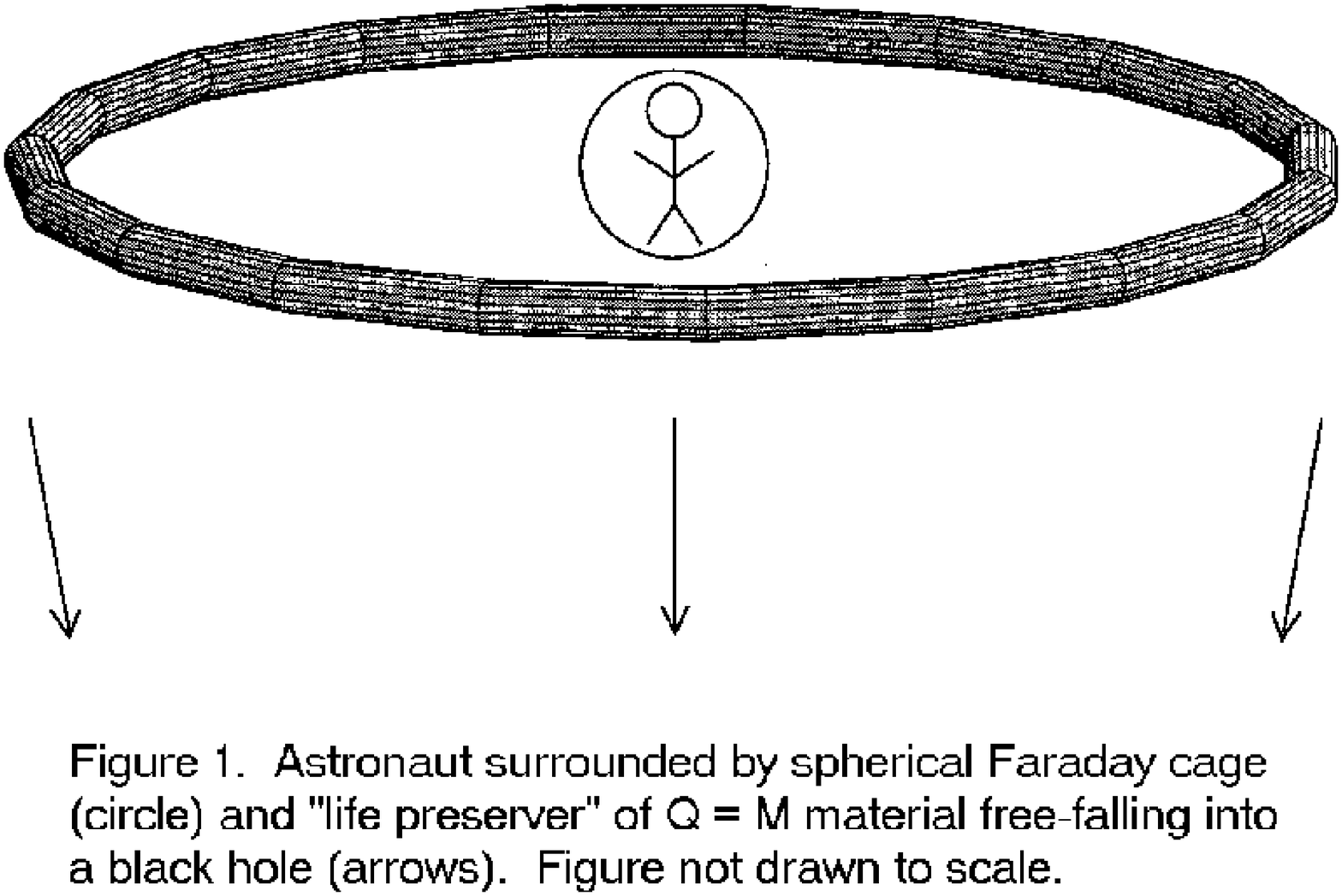}
\end{figure}

A thin ring of radius $R$ exerts an acceleration $a_z$ along the $z$-axis:
$a_{z}(z) = -M_{R}z/(z^2 + R^2)^{3/2}$.  Consider the ring to be a torus of 
major radius $R$ with cross sectional width $4\delta$ and height $4\epsilon$ 
where $\delta,\epsilon \ll R$.  It may be approximated as four thin rings, 
each of mass $M_R/4$, located at 
$x^2 + y^2 = (R \pm \delta)^2, z = \pm \epsilon$.  This torus produces an 
acceleration in the $z$ direction as a function of $z$ along the $z$-axis 
(for $z \ll R$) of approximately: 
\beq
a_z(z) \approx -\frac{M_R z}{R^3} \left ( 1 - \frac{3}{2}\frac{z^2}{R^2} + 6 \frac{\delta^2}{R^2} - \frac{9}{2}\frac{\epsilon^2}{R^2} \right ).
\eeq
 The first term in parentheses exactly cancels the tidal acceleration from the
 hole leaving a much smaller net acceleration
\beq
  a_{z, net}(z) \approx -\frac{M_R z}{R^3} \left ( - \frac{3}{2}\frac{z^2}{R^2} + 6 \frac{\delta^2}{R^2} - \frac{9}{2}\frac{\epsilon^2}{R^2} \right ).
\eeq

How might we make the ring neutrally stable against collapse under its own 
self-gravity?  One way would be to use $Q = M$ material where the electrostatic
 repulsion exactly cancels the gravitational attraction (Papapetrou [1], 
Majumdar [2], Bonner and Wickramasuriya [3]).  Surrounding yourself with an
 electrically conducting, spherical-shell Faraday cage of radius $h/2$ would 
protect you against the large electric fields, while still leaving you subject
 to the tidal gravitational field of the ring.  Since $m_e, m_p \ll e$, and 
$Q_R = M_R \ll h/2 \ll R$, the gravitational effects of introducing the 
Faraday cage and excluding the electric field inside are small.  And since 
$h/2 \ll R$ the electric quadrupole fields produced by the Faraday cage at the
 location of the ring are also fractionally small, of order $(h/2)^5/R^5$, 
and so do not affect its dynamics significantly.  During free-fall collapse 
$R \propto r, \delta \propto r, \epsilon \propto r^{-1/2}$ 
as each part of the ring follows a geodesic trajectory, and we require 
$-5 \mbox{g} \leq a_{z,net}(z) \leq 5 \mbox{g}$ for $-h/2 \leq z \leq h/2 $
 up until a time $t_c$ from the end.  You are neutrally stable at the center 
to first order because the tidal force of the black hole cancels that of the 
ring.  (For black holes larger than 7 solar masses, gradients in the strength 
of the hole's tidal field across your body are small relative to the maximum 
values of $a_{z,net}$ encountered in the situation described above.)	

Usefulness of $Q = M$ material is limited by electron-positron pair creation, 
whose rate is given by the relevant limit of Schwinger's formula 
(Schwinger [4]), valid when the electric field $E$ varies slowly over an
 electron Compton wavelength: 
\beq
\frac{dN}{dV dt} = \frac{\alpha E^2}{\pi^2 l_{p}^{2}} exp \left ( -\frac{\pi m_{e}^{2}}{| eE | l_{p}^{2}} \right ),
\eeq
 where, using geometrized units, $\alpha$ is the fine structure constant, 
$l_{p}$ is the Planck length, $e$ and $m_e$ are the electron charge and mass, 
and we integrate $dV$ over all space.  The ring has a total 
charge $Q_R = eN_0 = M_R$ and a discharge timescale $T_c = N_0/(dN/dt)$ since
 positrons will be repelled by the positively charged ring and escape while 
electrons will be trapped and discharge the ring.  We estimate $T_c$ by noting
 that, near $t_c$, if $a = 2\epsilon  = 2\delta \ll R$, the discharge timescale
 for the torus should be approximately equal to the discharge timescale for an
 infinite cylinder of radius $a$ and charge per unit length of 
$Q_R/ 2 \pi R$.  Integrating Schwinger's equation over all space then gives: 
\beq
T_c = \left ( \frac{\pi l_{p}^{2} R}{4 \alpha e M_R} \right )  \left [ x^4 \Gamma(-4,x) + \Gamma(0,x) \right ]^{-1},
\eeq
 where $\Gamma$ are incomplete gamma functions and 
$x = \frac{\pi^2 m_{e}^{2} R a}{e l_{p}^{2} M_R}$. As the ring 
shrinks, the discharge timescale becomes exponentially shorter.  Discharge is
 unimportant as long as $T_c \gg t_c$.  Using the above equations we establish
 an optimal solution where at the endpoint of usefulness, 
$t_c = 3.46 \times 10^{-3}$ sec, $R = 28.47$ m, 
$a = 2\epsilon = 2\delta  = 1.8$ m,
 $M_R = 2.145 \times 10^{-6} M_{\mbox{\scriptsize{earth}}}$, and 
$T_c = 1.3$ sec $\gg t_c$, with 50\% of the pairs being created near the 
surface of the torus (within $\pm3\%$ of the radius $a = 1.8$ m, the 
Faraday cage making a negligible correction to the calculation). 

Thus, the ring effectively counters black hole tides of 6760 g's across your 
body, allowing you to live longer and be tortured for only 
1/26$^{\mbox{\scriptsize{th}}}$ as long as before.  If you were killed in
 0.0904 seconds, pain signals starting at your waist (where you were being 
pulled apart) traveling along the fast pathways (6 - 30 m/sec) --  those 
telling you something is happening and where -- would not have time to arrive,
 so it would not ``hurt'' much.  By contrast, being killed in 
$3.46 \times 10^{-3}$ sec, you really wouldn't know what hit you.  In this 
example the number of pairs created prior to $t_c$ is large, of order 
$1.5 \times 10^{25}$, and the Faraday cage would have to protect you.  However,
 if $M_R = 9.451 \times 10^{-7}$ M$_{\mbox{\scriptsize{earth}}}$, 
$a = 2 \epsilon = 2 \delta = 1.8$ m, 
$R = 24.16$ m, at $t_c = 4.08 \times 10^{-3}$ sec, the
pair creation rate would be lowered so that, on average, no pairs would be 
created prior to $t_c$.

Finally, we would note that instead of falling in feet first, a better posture
 would be to lie in fetal position with the line connecting your shoulders
 pointing toward the black hole, thus squeezing your body into a cylinder 18 
inches high and 36 inches in diameter, taking advantage of the fact that the
 tidal acceleration is smaller by a factor of 2 in the $x,y$ directions than
in the $z$ direction, giving $t_c = 0.0455$ sec.  A massive ring as described 
above (at $t_c$ with $R = 7.242$ m, $a =  2 \epsilon = 2 \delta = 0.4572$ m, 
$M_R = 1.393 \times 10^{-7}$ M$_{\mbox{\scriptsize{earth}}}$ allows you to 
survive unscathed down to $t_c = 1.75 \times 10^{-3}$ sec.

It is interesting that, using only normal matter, we may in principle 
counteract tidal forces encountered in extreme situations.  This might also 
find application in trips near neutron stars or small black holes (without 
falling in) where an adjustable-radius, actively-oriented life preserver 
might enable you to venture closer than would otherwise have been the case
and still return safely home from the adventure.  When entering a region
of unknown tidal forces one might want to take along a large spherical
shell of $Q = M$ material.  Then, as needed, rings or small spherical 
masses could be drawn in from the shell to smaller radii to counteract
the vacuum tidal fields encountered.  A general tidal field produced
by an arbitrary distribution of distant matter in the Newtonian limit
could be countered by a series of rings, each ring countering each
point mass in the distant mass distribution.  Strong tidal forces due
to a gravitational wave (moving along the $z$ axis) could be 
countered by pulling in six small spherical masses on the $x$ and $y$
axes out of phase with the gravitational wave.  For the case where 
you fall into an unperturbed Kerr black hole metric along the
rotation axis to take advantage of the axisymmetry, you would need
to pull in a ring at first as in the Schwarzschild case, but would need
to manually adjust its radius as a function of time, finally 
enlarging it to large radius and replacing it with two small
equidistant spherical masses drawn in along the previous ring axis of
symmetry upon entering the region where $r < (3)^{1/2} a$, where the
tide from the black hole changes sign.  As you approach the inner
Cauchy horizon at $r_{\_}$, however, other effects would have to be
considered, such as infalling photons and gravitons and black hole
evaporation.  Quantum gravity effects would also have to be 
considered, as well as limits on the magnitude of the tidal forces
you can counter as we have discussed Schwarzschild case.  Still,
the results presented here prompt one to ask whether masses 
carried along might allow you to survive longer than would 
otherwise have been possible.

\acknowledgments{We thank Li-Xin Li and Jeremy Goodman for helpful comments.
JRG acknowledges support from his NSF grant AST-9900772.}



\begin{thebibliography}{99}

\bibitem{pap}
A. Papapetrou, Proc. R. Irish Acad. {\bf{51}}, 191 (1947).
\bibitem{maj}
S. D. Majumdar, Phys. Rev. {\bf{72}}, 390 (1947).
\bibitem{bon}
W. B. Bonner and S. B. P. Wickramasuriya, MNRAS {\bf{170}}, 643 (1975).
\bibitem{schw}
J. Schwinger, Phys. Rev. {\bf{82}}, 664 (1951).

\end{thebibliography}
\end{document}